\documentclass[fleqn,usenatbib]{mnras}

\usepackage{newtxtext,newtxmath}
\usepackage{threeparttable}
\usepackage{ulem}
\usepackage[T1]{fontenc}

\DeclareRobustCommand{\VAN}[3]{#2}
\let\VANthebibliography\thebibliography
\def\thebibliography{\DeclareRobustCommand{\VAN}[3]{##3}\VANthebibliography}

\usepackage{graphicx}	% Including figure files
\usepackage{amsmath}	% Advanced maths commands

\usepackage{amssymb}	% Extra maths symbols
\title[Broadband X-ray study of CD Ind]{A broadband X-ray study of the asynchronous polar: CD Ind}

\author[Dutta and Rana]{
Anirban Dutta,$^{1}$\thanks{E-mail: anirband@rri.res.in}
Vikram Rana$^{1}$
\\
$^{1}$Astronomy and Astrophysics Group, Raman Research Institute, Bangalore, India\\
}

\date{Accepted XXX. Received YYY; in original form ZZZ}

\pubyear{2015}

\begin{document}
\label{firstpage}
\pagerange{\pageref{firstpage}--\pageref{lastpage}}
\maketitle

\begin{abstract}
%This is a simple template for authors to write new MNRAS papers. The abstract should briefly describe the aims, methods, and main results of the paper. It should be a single paragraph not more than 250 words (200 words for Letters). No references should appear in the abstract.

A simultaneous broadband analysis of X-ray data obtained with XMM-Newton and NuSTAR observatories for the asynchronous polar source CD Ind is presented. The spin folded lightcurve in soft 0.3-3.0 keV band shows single broad hump-like structure superimposed with occasional narrow dips, indicating a single-pole accretion model with a complex intrinsic absorber. Lack of strong modulation in folded lightcurve above 3 keV reveals that emission from corresponding zone of post-shock region (PSR) remains in view throughout the spin phase. The broadband spectrum is modelled with a three-component absorbed plasma emission model and absorbed isobaric cooling flow model, both of which fit the data well with similar statistical significance. Presence of partial covering absorber is evident in the spectra with equivalent column density $\sim7\times10^{22}\;\text{cm}^{-2}$ and a covering fraction of $\sim 25\%$. Strong ionised oxygen K$_{\alpha}$ line emission is detected in the spectra. We notice spectral variability during spin phase 0.75-1.05, when there is a considerable increase in column density of overall absorber (from $\sim 1 \times 10^{20}\;\text{cm}^{-2}$ to $\sim 9 \times 10^{20}\;\text{cm}^{-2}$). We required at least three plasma temperatures to describe the multi-temperature nature of the PSR. The shock temperature $\sim 43.3_{-3.4}^{+3.8}$ keV, represented by the upper temperature of the cooling flow model, implies a white dwarf mass of $\sim 0.87^{+0.04}_{-0.03}\;M_{\odot}$. The iron K$_{\alpha}$ line complex shows a strong He-like and a weak neutral fluorescence line. We could not unambiguously detect the presence of Compton reflection in the spectra, which is probably very small and signifying a tall shock height.
\end{abstract}


%%%%%%%%%%%%%%%%%%%%%%%%%%%%%%%%%%%%%%%%%%%%%%%%%%

%%%%%%%%%%%%%%%%% BODY OF PAPER %%%%%%%%%%%%%%%%%%

\section{Introduction}

Magnetic cataclysmic variables (mCVs) are a type of binary systems where the accreting material, coming from the Roche-lobe filling secondary star (typically late type main sequence star), falls on the primary white dwarf (WD) with supersonic velocities via the magnetic field lines of the WD \citep{Warner_1995, Hellier_2001}. Asynchronous polars(APs) are the type of mCVs which have spin periods ($\omega$) different than orbital periods ($\Omega$), but very close to each other (with asynchronicity,  $\frac{|P_{\Omega}-P_{\omega}|}{P_{\Omega}} \lesssim3\%$). Though intermediate polars(IPs) also have differing spin and orbital periods, the degree of asynchronicity  is much higher (for most IPs $\gtrsim90\%$,  eg. \citet{Bernardini_2012, deMartino_2020}). The magnetic field strength of polars are high enough ($\gtrsim10MG$), so that the field lines of the primary extend up to secondary and intertwines with field lines of the secondary, thus causing magnetic locking and resulting perfect synchronicity. IPs have weaker magnetic field ($\sim1-10$MG) and unable to have magnetic locking, thereby posses high level of asynchronicity. APs have comparable magnetic fields as polars, and their spin periods gradually evolve towards the orbital period. The origin of asynchronicity in APs is still under debate. For one AP : V1500 Cyg, nova eruption \citep{Honda_1975, Stockman_1988, Schmidt_1995} is theorised to be a reason behind asynchronicity based on detection of nova shell around it. However, nova shells remain undetected around other APs \citep{Pagnotta_2016}. The WD in AP rotates over a beat period ($|1/P_{\omega}-1/P_{\Omega}|^{-1}$) with respect to the secondary. Thus it provides insight about how the accretion properties vary in different intervals of beat cycle.

In magnetic CVs including APs, the supersonic accretion flow, channelled via magnetic field lines, form strong shock over WD surface near the pole region and produces X-rays. The material cools down in the post-shock region (PSR) primarily via bremsstrahlung emission in X-ray, and cyclotron emission in optical wavelength \citep{Cropper_1990}. The PSR of the accretion column thus represents multi-temperature zones of ionised plasma as it cools down at the bottom of the column i.e. the WD photosphere \citep{Aizu_1973}. The observed X-rays from the system carries the information about its interaction with the WD surface, accretion column, accretion disk or any other galactic intervening medium. A part of the X-rays from PSR can raise WD surface temperature and produce blackbody emission in very soft X-rays (few eVs to tens of eVs) to extreme UVs. A part of the interacting X-rays will undergo photoelectric absorption, the effect which is more pronounced as a reduction in soft X-rays (few hundreds of eVs to few keVs). There will also be fluorescence line emission, most notably the neutral Fe $K_{\alpha}$ line at 6.4 keV, followed by photoelectric absorption by WD surface.  A part of downward emitting hard X-rays can undergo Compton reflection by the WD surface which will present itself as an excess (a hump like feature) in 10-30 keV \citep{George_1990}. This last process is highly dependent on the parameters related to geometry of the accretion, like shock height and viewing angle of reflecting site. Also the strength of the reflection amplitude is correlated with the strength of the 6.4 keV line as they are believed to be originated from the same region of WD.

Our target CD Ind (also known as EUVE J2115-586, RX J2115-5840) is one of the asynchronous polars (other notable few V1432 Aql, BY Cam and V1500 Cyg) which has been first identified as polars by \citet{Craig_1996, Vennes_1996} via spectroscopic studies and estimating the magnetic field strength. \citet{Schwope_1997} updated the definition of CD Ind as an AP by intensive polarimetric study and signature of pole-switching was observed over a beat cycle. They also reported a magnetic field strength of $11\pm2$MG. \citet{Ramsay_1999, Ramsay_2000} stated  that the acctreting material follows same set of field lines over the full beat cycle, i.e. the materials travel around the azimuth of WD to connect to those field lines over certain beat phases and argued about the complex magnetic field structure with non-dipolar field geometry of the WD, where one pole is sufficiently stronger than other. \citet{Ramsay_2000} also performed X-ray analysis using RXTE PCA data, taken over several days of a beat cycle and found that the hard X-ray spectra in 4-15 keV produces similar spectral parameters for both the accreting poles, whichever is active for accretion at different beat phases. Involving a far ultraviolet spectroscopic study, \citet{Araujo-Betancor_2005} reported a galactic column density of $1\pm0.5 \times 10^{19}\; cm^{-2}$ by measuring the absorption of a narrow interstellar Ly$\alpha$ line. \citet{Myers_2017} revisited the source with extensive photometric campaign extended over a duration of 9 years and defined the spin and orbital periods along with the rate of change of spin period ($\dot{P_{\omega}}$). Later \citet{Littlefield_2019} used continuous TESS photometric data of 28 days and redefined the Myer's identified period with $P_{\Omega}=6720s$ and $P_{\omega}=6648s$. They also claimed that $\dot{P_{\omega}}$ is half of what Myer's identified and updated the resynchronisation time scale ($\tau=|\frac{P_{\omega}-P{\Omega}}{\dot{P_{\omega}}}|$) to be 13000 years, making it one of the slowest achiever of synchronicity. \citet{Littlefield_2019} reaffirmed that one accretion region is continuously visible during accretion over a spin phase, when the other pole undergoes self-eclipse and each pole accretes for nearly half of the beat cycle ($\sim7.3$ days). Using the TESS data, \citet{Hakala_2019} studied the changes in accretion stream trajectory on to the two pole. The following year, also employing the TESS data, \citet{Mason_2020} discussed the possible accretion scenario where four alternating and oppositely positioned accretion regions are present, with one accretion region being always in view. \citet{Sobolev_2021} performed magnetohydrodynamic simulation of the flow structure under the assumption of shifted dipole configuration and predicted significant changes in flow structure depending on the pole-switching.

In this work, we present the study of the asynchronous polar source, CD Ind, for the first time, using the broadband X-ray data, obtained from  XMM-Newton and NuSTAR telescope. We organise our paper as follows. In the next section (Sect.~\ref{sec:ObsDataRed}) we present the observations used and data reduction. Section ~\ref{sec:AnalysisResults} contains the results from the timing and spectral analysis of the source. In section ~\ref{sec:discussion} we discuss the results obtained from previous sections. The concluding section (sect. ~\ref{sec:summary}) describes the summary of the work.

\section{Observations and data reduction}
\label{sec:ObsDataRed}

CD Ind was observed simultaneously with XMM-Newton \citep{Jansen_2001} and NuSTAR telescopes \citep{Harrison_2013} as a part of our proposal to perform a detailed broadband X-ray spectral study of APs. The hard X-ray imaging telescope NuSTAR is capable of extending our understanding till 79 keV with high sensitivity. Simultaneous observation with XMM-Newton, having excellent energy resolution, empowers us to probe the soft energy part till 0.3 keV. So, the availability of simultaneous broadband data provides us the superior opportunity to characterise the spectrum by accounting the absorption in lower end as well as probing the reflection in upper end, thereby constraining the multi-temperature continuum from PSR. XMM-Newton telescope observed (Observation ID: 0870800101) the source for $\sim 36.8 \;\text{ks}$ on source time, starting at 2020-11-09 T12:32:16 and NuSTAR observed (Observation ID: 30601018002) for $\sim 56.5\;\text{ks}$, starting at 2020-11-09 T12:11:09. The high resolution reflection grating spectrometer, RGS \citep{denHerder_2001} on-board XMM, covering 0.35-2.5 keV energy band, can resolve the prominent emission lines present in the source.

\subsection{NuSTAR}

The two focusing imaging telescope modules of NuSTAR, namely FPMA and FPMB, can bring the hard X-rays (3.0-79.0 keV) to its focus and record with high sensitivity. We have selected a circular source region of 40 arcsec radius centering the source, and a circular source free region of 80 arcsec radius as the background from the same detector. 
We have used NuSTARDAS version 2.1.1 for data reduction. The latest calibration files are used (v20210701).  \texttt{nupipeline} is used to obtain the cleaned event files using default screening criteria. The \texttt{nuproducts} command has been used to produce final science data products like lightcurve and spectrum files and necessary detector response files. We have performed the barycentric correction during product extraction. We have rebinned the NuSTAR spectra with minimum 25 counts in each bin using \texttt{grppha} to utilize the $\chi^2$ minimization for spectral fitting. 

\subsection{XMM-Newton}

XMM-Newton observation of the source was taken in large window mode using thin filter for both the PN \citep{Struder_2001} and MOS detectors \citep{Turner_2001} of the European Photon Imaging Camera Instrument (EPIC). We have used XMMSAS v19.1.1 for the data reduction. The calibration files used, are obtained from SAS current calibration files repository, latest at the time of analysis \footnote{\url{https://www.cosmos.esa.int/web/xmm-newton/current-calibration-files}}. We have followed the SAS analysis thread \footnote{\url{https://www.cosmos.esa.int/web/xmm-newton/sas-threads}} for data reduction. We have used SAS tools \texttt{epproc} and \texttt{emproc} to produce calibrated event files. Our data are contaminated heavily by high background flair due to XMM-Newton's highly elongated eccentric orbit. The flaring is prominent during later part of observation in PN and MOS data. To get rid of this flaring, we discarded the data using time selection criteria \texttt{TIME<721326500} in our good time interval (GTI). Unfortunately, this aggressive but essential filtering leaves us with only initial $\sim 11.2$ ks and $\sim13.1$ ks of data from PN and MOS respectively. We have also checked for pile-up using \texttt{epatplot} tool but did not find any significant presence of it. The flaring free event files are then used for science products extraction with barycentric correction. We have chosen a circular source region with 25 arcsec radius centering the source, and a circular background region with 50 arcsec radius from the same CCD to extract our final lightcurve, spectrum and detector response files. The spectra have been rebinned with \texttt{spacegroup} tool to minimum 25 counts for using $\chi^2$ statistic to test goodness of spectral fit.

For RGS data extraction we have used \texttt{rgsproc} tool. Though we found that RGS data are not contaminated to that hefty extent as that of EPIC, yet background flaring peaks are present, for which we have used rate selection criteria \texttt{RATE<=0.125} in the corresponding good time interval. We managed to get $\sim30.9$ ks of the exposure time for the spectra obtained from RGS detectors for our analysis. Spectra were rebinned with minimum 25 counts like before.

\section{Data Analysis and Results}
\label{sec:AnalysisResults}

\subsection{Timing Analysis}

\begin{figure}
	\includegraphics[width=\columnwidth, trim = {0.7cm 0cm 0.1cm 0cm}, clip]{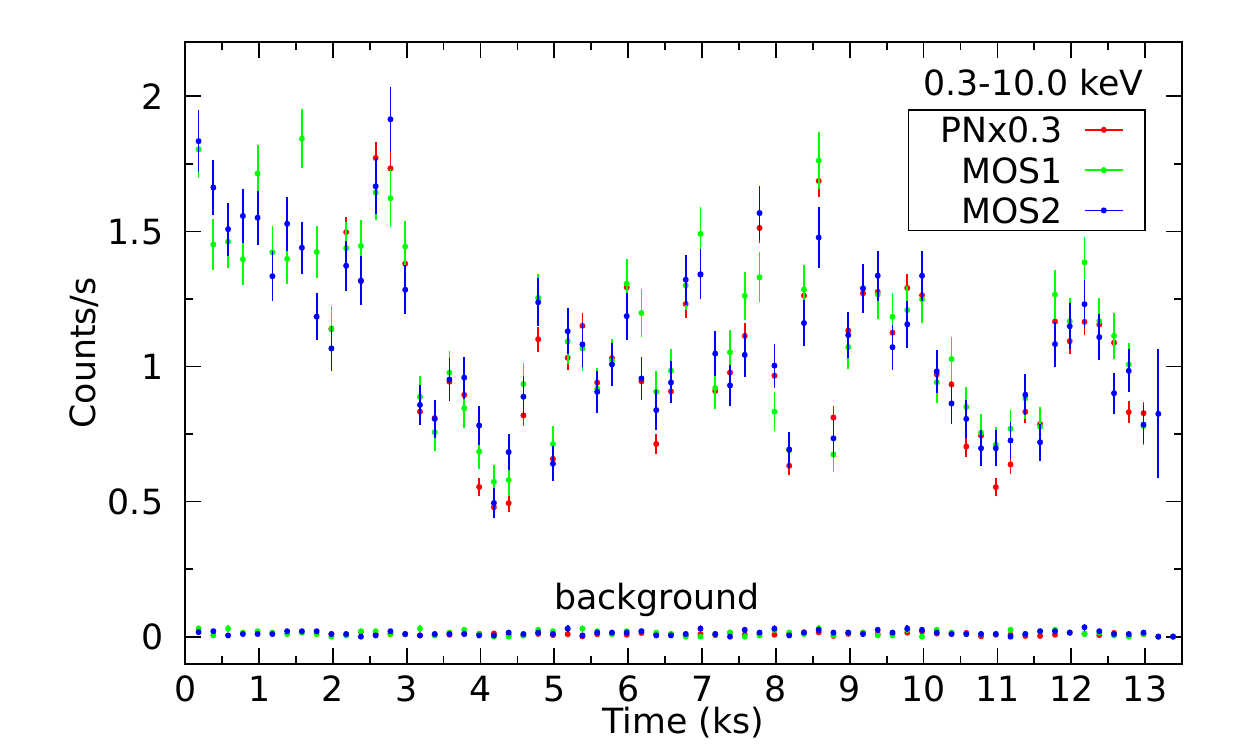}
	\includegraphics[width=\columnwidth, trim = {0.7cm 0cm 0.1cm 0cm}, clip]{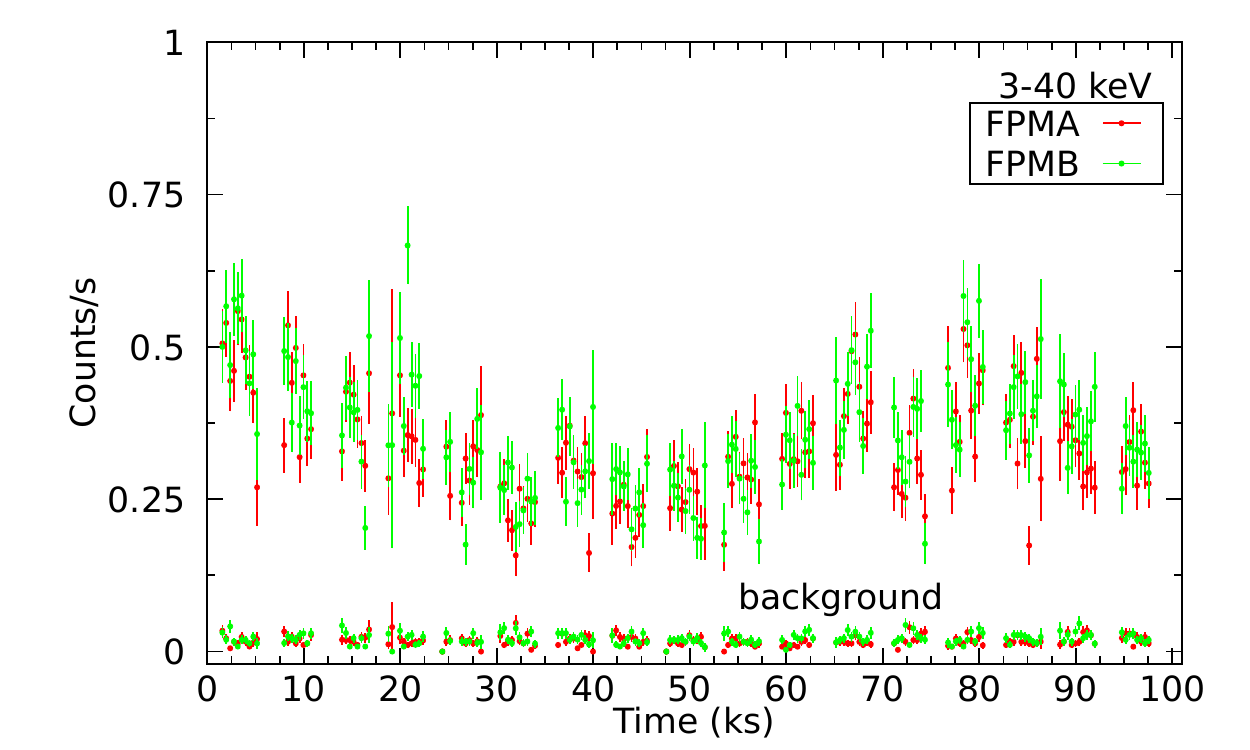}
	\includegraphics[width=\columnwidth, trim = {0.7cm 0.1cm 0.1cm 0.1cm}, clip]{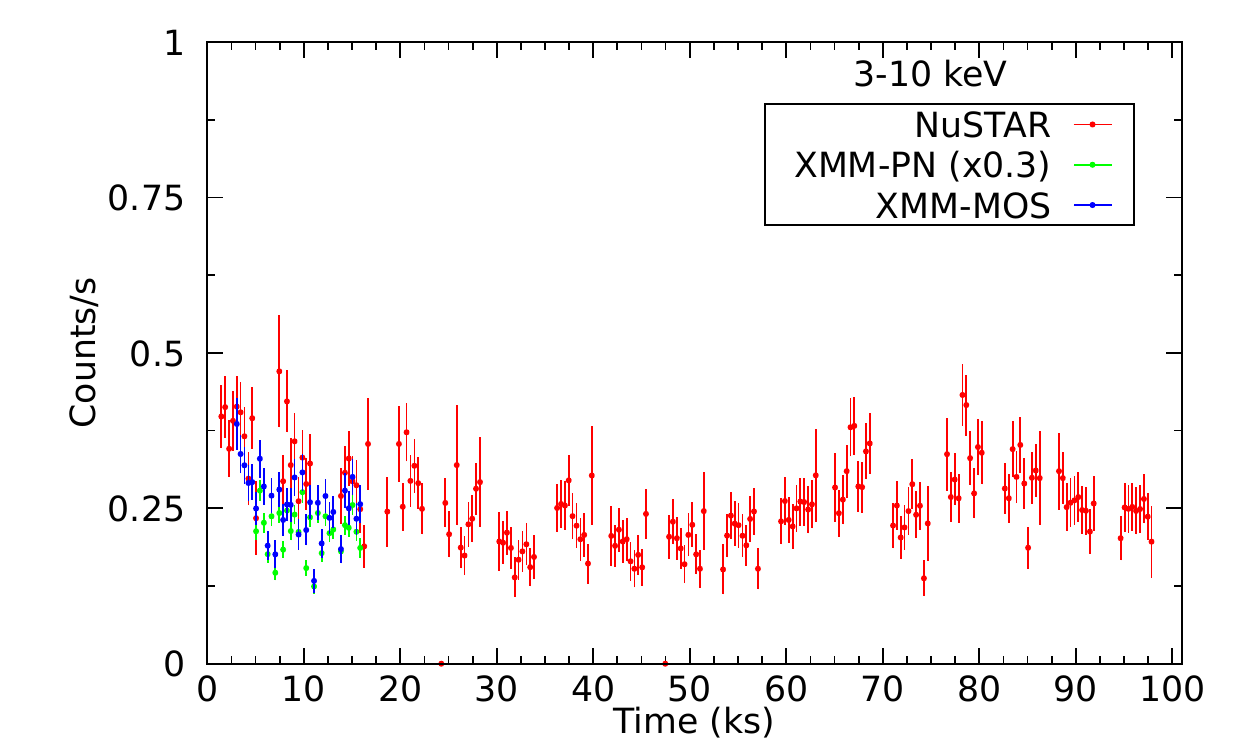}

    \caption{Background subtracted clean lightcurves from XMM-EPIC in 0.3-10.0 keV (\textcolor{blue}{top}) and NuSTAR in 3-40 keV (\textcolor{blue}{middle}) . Background lightcurves are also shown in those panels for reference. In the \textcolor{blue}{bottom} panel, cleaned lightcurves are shown in the common 3-10 keV band where both telescopes have coverage. The lightcurves from both the modules of NuSTAR are co-added and averaged to improve statistics. Same is done for XMM-Newton EPIC-MOS. The start time is chosen to be NuSTAR observation start time. The PN counts/s is scaled by 0.3 in top and bottom panel for comparison. The bin size is 200s in top panel and 400s for middle and bottom panel.}
    \label{fig:cleanlight}
\end{figure}

The data obtained from NuSTAR observation is $\sim96$ ks including the actual on-source time of $\sim56.5$ ks, gaps due to earth occultation and South Atlantic Anomaly (SAA) passage. These gaps result in difficulties to find the exact periods of the system with high precision and to distinguish between the spin ($6648s$) and orbital periods ($6720s$) which are closely spaced. However, given the total duration of NuSTAR data, we can get roughly $\sim14$ cycles to probe the timing properties of the system. On the other hand, flaring corrected EPIC data cover only $\lesssim2$ cycles. For spin and orbital period, we have followed \cite{Littlefield_2019} who defined the periods based on nearly 28-days long continuous TESS data.
The background subtracted cleaned lightcurves from both observatories have been plotted in the Figure~\ref{fig:cleanlight}.

We performed power spectral analysis on background subtracted PN and FPMA lightcurves that showed broad peaks at $6286\pm917$s and $6545\pm111s$ respectively. These values agree with the literature values of rotational periods of the system, but our data could not resolve the spin and orbital periods.

\subsubsection{Spin Folded Lightcurves} 

To fold the lightcurve based on the spin period, we have used the ephemeris $T(BJD)=2458326.46492(17)+0.0769522(11) \times E$ following \cite{Littlefield_2019}. They updated the derivative of spin period ($\dot{P}=+1.75\times10^{-10}$) from the value given by \cite{Myers_2017} which was two times faster. The time when our observation was made (BJD=2459163.00774), the spin period has changed only by $\sim0.0126$s i.e $\sim0.0002\%$ from the reported value in \cite{Littlefield_2019}.

\begin{figure}
	\includegraphics[width=\columnwidth, trim = {0cm 0cm 0cm 1cm}, clip]{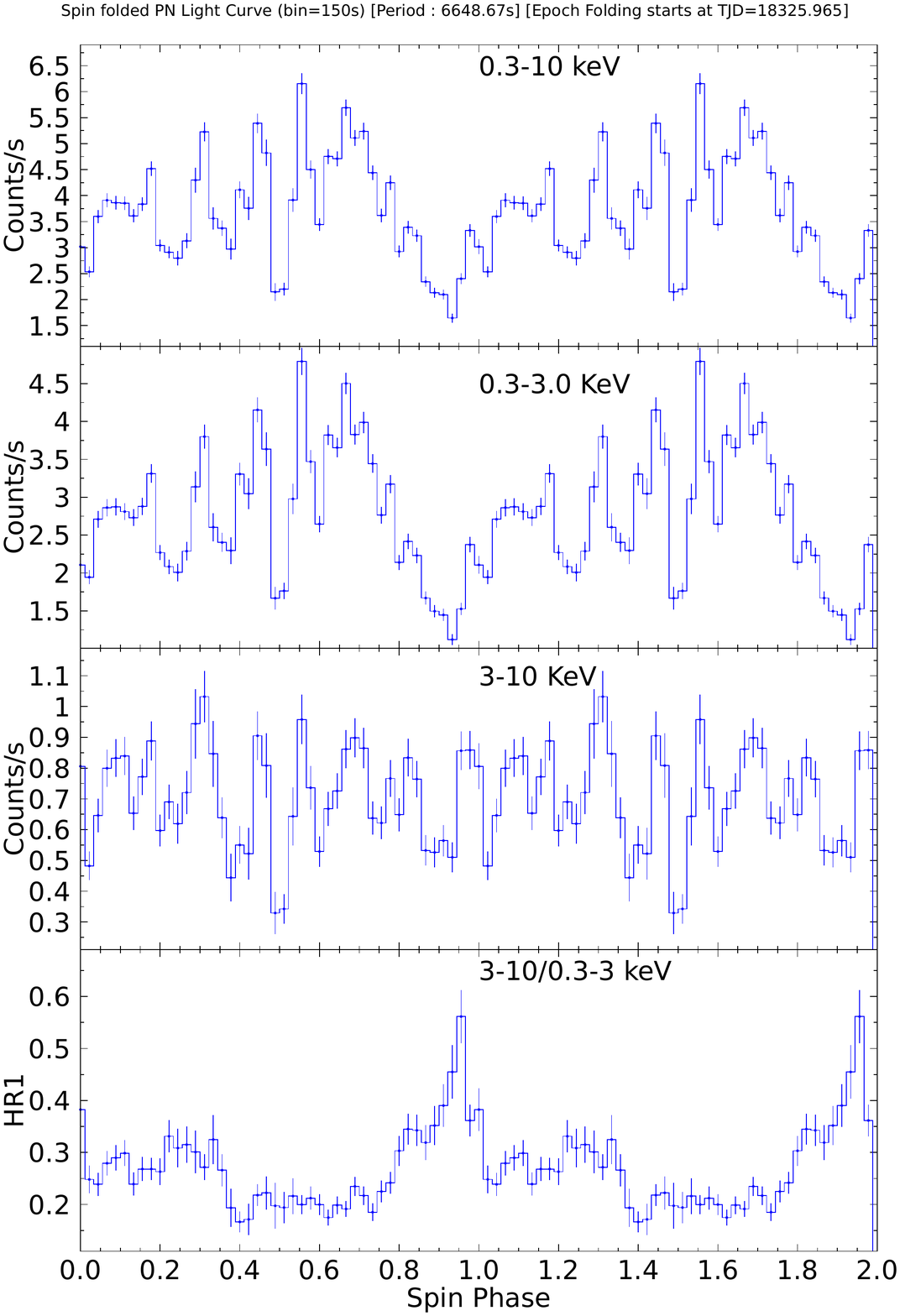}

    \caption{Spin folded lightcurves in different energy bands along with hardness ratio using XMM-EPIC PN data. Bin size in each panel is  $\sim150s$}
    \label{fig:spinfoldPN}
\end{figure}

We have shown the background subtracted spin folded XMM-Newton PN lightcurves in Fig. \ref{fig:spinfoldPN} for different energy bands (0.3-10 keV, 0.3-3.0 keV, 3.0-10.0 keV).
The soft X-ray band (0.3-3.0 keV) exhibit a strong pulse fraction (PF) of modulation, $62\pm2\%$, using the definition PF $=(I_{max}-I_{min})/(I_{max}+I_{min})$ where $I$ denotes the count rate.  It shows a single broad hump like structure with occasional narrow dips in between. However, it is difficult to comment about all the dips individually due to the limitation of the data, which cover $\lesssim2$ cycles. In 3-10 keV band of PN, the broad hump-like profile is not visible, and the count rate fluctuates around a mean value of 0.673 counts/s. The hardness ratio-1 plot ($HR1=I_{3-10\;\text{keV}}/I_{0.3-3.0\;\text{keV}}$) in the bottom panel of Fig. \ref{fig:spinfoldPN} shows a strong peak in phase $0.75-1.05$, denoting a significant spectral variability during that phase.

\begin{figure}
	\includegraphics[width=\columnwidth, trim = {0cm 0cm 0cm 1cm}, clip]{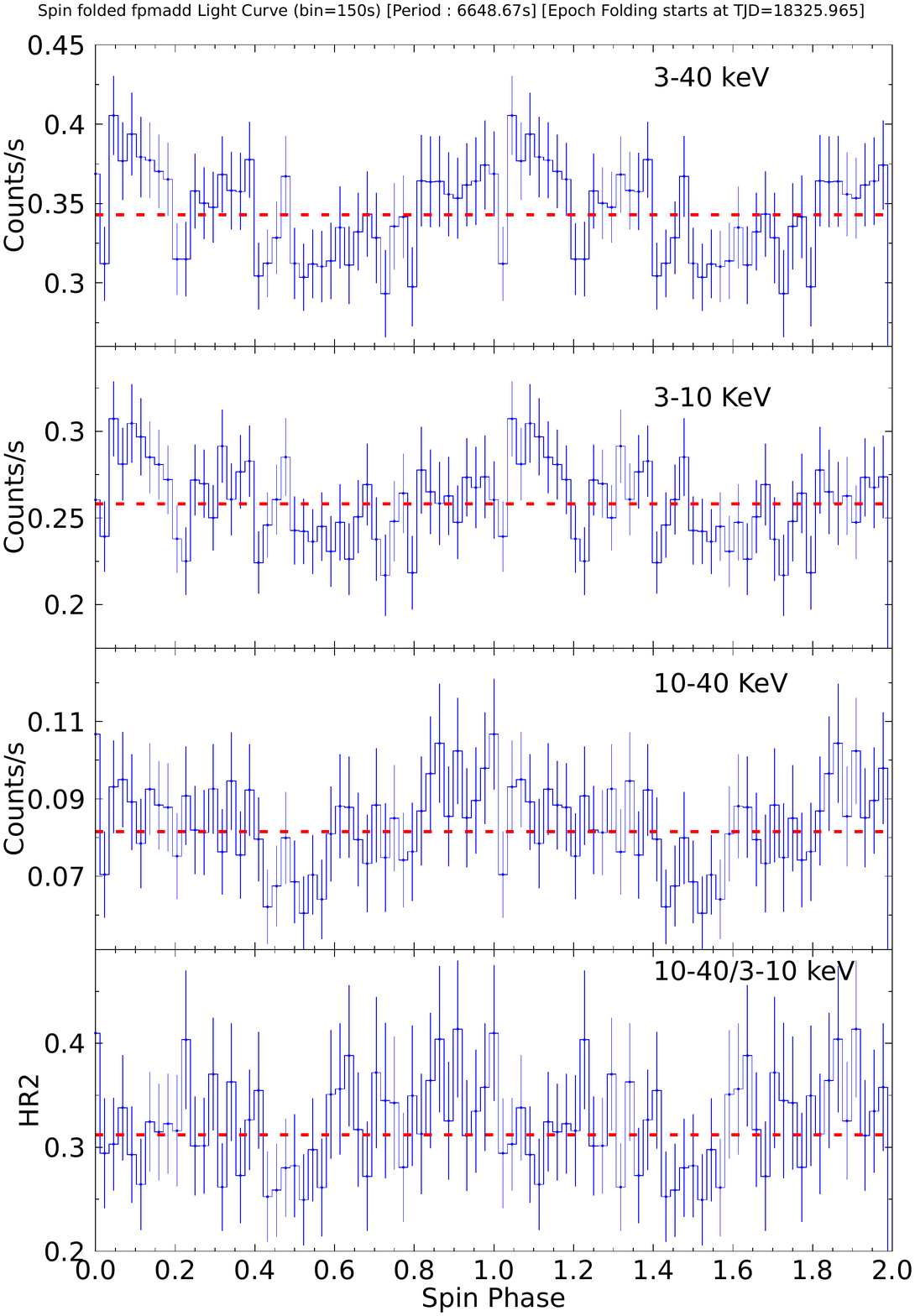}

    \caption{Spin folded lightcurves in different energy bands along with hardness ratio using co-added and averaged lightcurves from both the FPMA and FPMB modules of NuSTAR. Bin size in each panel is $\sim150s$. The red dotted line represents mean counts rate by fitting a \texttt{constant} model}
    \label{fig:spinfoldFPMA}
\end{figure}

Background subtracted spin folded NuSTAR lightcurves in different energy bands (3-40 keV, 3-10 keV, 10-40 keV) are plotted in Fig. \ref{fig:spinfoldFPMA}. We have chosen 3-40 keV band from NuSTAR, since the background starts dominating beyond 40 keV. The folded lightcurve in 3-40 keV band in top panel of Figure \ref{fig:spinfoldFPMA} statistically represents a flattened profile with an average value $\sim 0.343$ counts/s.  We observe a similar flattened profile with fluctuating count rates in both the constituent energy bands, 3-10 keV and 10-40 keV. For the profile in 3-10 keV and 10-40 keV band, a \texttt{constant} model can fit reasonably well with an amplitude of $\sim0.258 \pm 0.004$ counts/s ($\chi^2_{\nu}(DOF)=1.149(87)$) and  $0.081\pm0.002$ counts/s ($\chi^2_{\nu}(DOF)=0.873(87)$) respectively. Also, the hardness ratio-2 ($HR2=I_{10-40\;\text{keV}}/I_{3.0-10.0\;\text{keV}}$) in the bottom panel of Figure \ref{fig:spinfoldFPMA}, show a flat profile with an average value $\sim0.312$, indicating there is no significant spectral variability above 3 keV over the full spin period.

The orbit folded lightcurves, using period $6720$s \citep{Littlefield_2019}, produce similar pattern in all energy bands as well as the hardness ratios, as observed in the spin folded lightcurves.

\subsection{Spectral Analysis}

We have used X-ray spectral fitting package, XSPEC \citep{arnaud_1996} version 12.12.0, to analyse the spectra. The spectral models used in this paper are available in the package. The errors on the spectral parameter values are quoted with 90\% confidence throughout the paper. Abundance table in our analysis was set after Wilms \& McCray abundance table %\footnotemark[\getrefnumber{footsetting1}]
\citep{Wilms_2000} along with photoelectric absorption cross section defined after \cite{Verner_1996} 
%\footnotemark[\getrefnumber{footsetting1}].

\subsubsection{Phenomenological fit to NuSTAR spectra}

We have first utilised only the NuSTAR data for our phenomenological fit to get the idea about shock temperature and reflection amplitude. Both the FPMA and FPMB spectra (3.0-40.0 keV) were fitted simultaneously to improve the signal to noise ratio. The equivalent column density of galactic absorber (modelled after \texttt{tbabs} ) was fixed at $10^{19}\; \text{cm}^{-2}$ \citep{Araujo-Betancor_2005} as NuSTAR data can not constrain it.
Ionised plasma emission model \texttt{mekal} has been used for modelling the emission from the post-shock region. The switch parameter for the emission model was set at 2 to determine the spectrum based on updated line emission code \texttt{AtomDB v3.0.9}
%\footnote{\label{footsetting1} Same setting for subsequent spectral analysis all throughout this paper}. 
To incorporate the multi-temperature nature of the post-shock region of the accretion column, we have used two \texttt{mekal} components, where the upper temperature gives indication of the shock temperature. 

The iron line complex (neutral fluorescence at 6.4 keV, He-like at $\sim6.7$keV and H-like at $\sim6.97$keV) is not separately resolved by NuSTAR, and was modelled using a \texttt{gaussian} component. The combined model (including absorption, emission and gaussian) produced an upper temperature $ 30.7_{-6.9}^{+19.7}\; \text{keV}$ with a $\chi^2(DOF)=411(459)$.  Next we convolved the \texttt{reflect} model \citep{Magdziraz_1995} with the emission components, to find out the effect of the Compton reflection which should manifest itself as an excess or hump in $\sim10-30$ keV energy range. We have kept the parameters of \texttt{reflect} other than reflection amplitude fixed at their default values (viewing angle set at $\mu=0.45$, the abundance parameters were linked between reflection and emission component and were kept frozen at 1). The fit statistics didn't have any significant change ($\chi^2(DOF)=409(458)$ i.e $\Delta\chi^2 \sim 2$ for 1 less DOF), indicating redundancy of \texttt{ reflection} component in the fitting. However, NuSTAR-only fitting can not detect the lower temperature of the PSR, as well as presence of any extra absorber, which affects the soft X-rays, mostly below NuSTAR's coverage. So, we need to incorporate the simultaneous data obtained from XMM-Newton for a global description of the spectra.

\subsubsection{Phenomenological fit to XMM-Newton spectra}

\begin{figure}
	\includegraphics[width=\columnwidth, trim = {0.5cm 0.4cm 3.6cm 2.7cm}, clip]{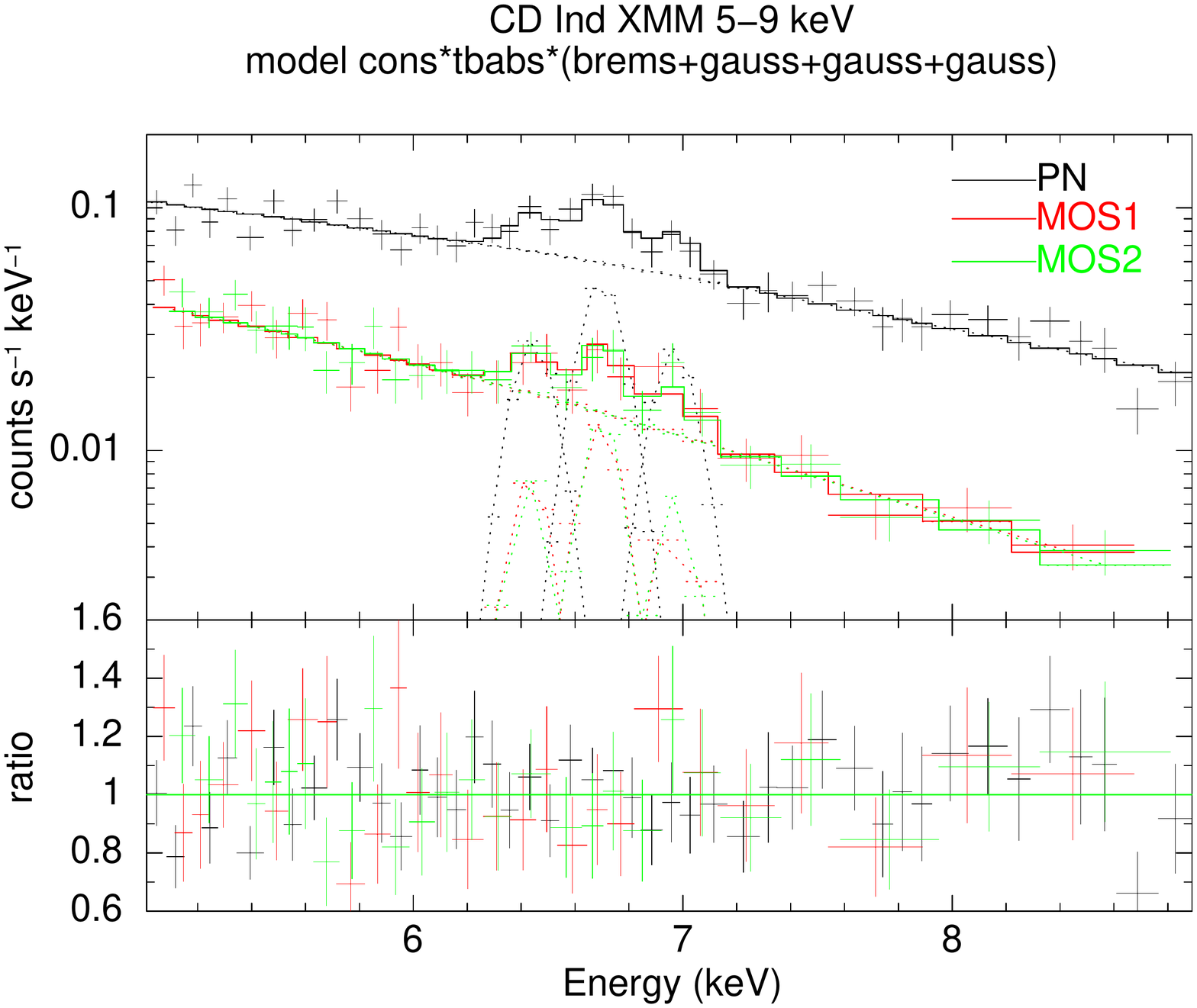}
    \caption{ Modelling the iron K$_{\alpha}$ line complex using an absorbed bremsstrahlung continuum along with three gaussian component in 5-9 keV. The \textcolor{blue}{top} panel shows the spectrum, and the \textcolor{blue}{bottom} panel shows ratio (data/model) plot.}
    \label{fig:ironfig}
\end{figure}

To build up the description of the soft X-ray part of the spectrum, we have looked into the XMM-Newton EPIC (0.3-10.0 keV) and RGS (0.45-2.0 keV) data. Owing to good spectral resolution of the EPIC, we can distinguish the Fe $K_{\alpha}$ line complex. In order to quantify the contribution of those three lines, we have modelled them after three \texttt{gaussian} on top of a thermal \texttt{bremsstrahlung} continuum with fixed galactic absorption ($10^{19}\;\text{cm}^{-2}$) using 5.0-9.0 keV data. The resultant best fit parameters are quoted in Table~\ref{tab:irontab} and corresponding spectral plot is shown in top panel of Fig.~\ref{fig:ironfig}. The width ($\sigma$) of all the \texttt{gaussian} components are consistent with the EPIC resolution limit ($\sim130$ eV at 6.5 keV), hence fixed at zero, thus indicating that the lines are narrow. The line centers appear at expected energies within error bar. Fluorescent line is the weakest (equivalent width $\sim71$ eV) whereas the He-like line is the strongest (equivalent width $\sim137$ eV) among the three lines.

\begin{table}
	\centering
	\caption{Probing Fe-K$_{\alpha}$ lines  (5.0-9.0 keV EPIC data)}
	\label{tab:irontab}
	\renewcommand{\arraystretch}{1.4}
\begin{tabular}{cccc}
\hline
Parameter          						& Unit                  & Value           \\ \hline
n$_{H}^{\dagger\;a}$       	    		& $10^{19}\;\; cm^{-2}$ & $1_{\mathit{fr}}$      \\ \hline
T$_{C}^{\dagger\;b}$					& keV                   & $32.8_{-14.5}^{+67.8}$ \\
N$_{C}^{\dagger\;c}$					& $10^{-3}$             & $2.24_{-0.08}^{+0.29}$ \\ \hline
LineE$_{\text{Neutral}}^{\dagger\;d}$  	& keV                   & $6.43_{-0.03}^{+0.04}$ \\
$\sigma_{\text{Neutral}}^{\dagger\;e}$ 	& eV                    & $0_{\mathit{fr}}$		 \\
eqw$_{\text{Neutral}}^{\dagger\;f}$    	& eV                    & $71_{-30}^{+37}$       \\
N$_{\text{Neutral}}^{\dagger\;g}$      	& $10^{-5}$             & $0.90_{-0.32}^{+0.32}$ \\ \hline
LineE$_{\text{He-like}}^{\dagger\;d}$  	& keV                   & $6.70_{-0.02}^{+0.02}$ \\
$\sigma_{\text{He-like}}^{\dagger\;e}$ 	& eV                    & $0_{\mathit{fr}}$		 \\
eqw$_{\text{He-like}}^{\dagger\;f}$    	& eV                    & $137_{-43}^{+46}$      \\
N$_{\text{He-like}}^{\dagger\;g}$      	& $10^{-5}$             & $1.67_{-0.37}^{+0.38}$ \\ \hline
LineE$_{\text{H-like}}^{\dagger\;d}$   	& keV                   & $6.97_{-0.03}^{+0.04}$ \\
$\sigma_{\text{H-like}}^{\dagger\;e}$  	& eV                    & $0_{\mathit{fr}}$	     \\
eqw$_{\text{H-like}}^{\dagger\;f}$     	& eV                    & $83_{-33}^{+46}$       \\
N$_{\text{H-like}}^{\dagger\;g}$       	& $10^{-5}$             & $0.93_{-0.35}^{+0.35}$ \\ \hline
$\chi^{2}(DOF)$				    		&                       & 81.40(89)              \\ \hline
$\chi^2_{\nu}$    			 			&                       & 0.9146                 \\ \hline
\end{tabular}

\begin{tablenotes}
      \small
      \item $\dagger\;a$ ISM column density
	  \item $\dagger\;b$ : Bremsstrahlung continuum temperature
	  \item $\dagger\;c$ : Normalisation of bremsstrahlung continuum
	  \item $\dagger\;d, e, f, g$ : Line energy, $\sigma$, equivalent width, and normalisation (in terms of photons cm$^{-2}$ s$^{-1}$) of the corresponding gaussian component respectively
	  \item * $\mathit{fr}$ denotes the parameter is fixed.

\end{tablenotes}

\end{table}

Next we have fitted the full EPIC spectra (PN, MOS1 and MOS2) in 0.3-10.0 keV to estimate the lower temperature and absorption parameters. A simple model like an absorbed single temperature ionised plasma emission (\texttt{mekal}) model  with gaussian component for narrow 6.4 keV line produced a mediocre fit with $\chi^2(DOF)=549(401)$ and a plasma temperature of $24.4_{-2.4}^{+2.6} \;\text{keV}$. But the fit has issues like extremely low value of column density of the absorber ($\sim10^{10}\;\text{cm}^{-2}$) unbounded at lower limit and with upper limit reaching  $\sim 2\times 10^{19}\;\text{cm}^{-2}$ (close to the galactic $n_{\text{H}}$ value of $10^{19}\;\text{cm}^{-2}$), and the excess around 0.6 keV and 1 keV indicating line emissions like oxygen and iron-L shell from low temperature plasma. Fixing the $n_{\text{H}}$ to $10^{19}\;\text{cm}^{-2}$ and adding one more plasma emission component resulted in a somewhat improved fit statistic  ($\chi^2(DOF)=527(400)$) with low temperature coming around $0.17_{-0.03}^{+0.02} \;\text{keV}$ representing an optically thin cold plasma. In order to evaluate the column density of any absorber present at the source, we included an photoelectric absorber model \texttt{phabs}, with $n_{\text{H,tb}}$ of galactic absorber \texttt{tbabs} fixed at $10^{19}\;\text{cm}^{-2}$ and found a better fit-stat ($\chi^2(DOF)=518(399)$) with $n_{\text{H,ph}}=1.18_{-0.67}^{+0.69}\times10^{20}\;\text{cm}^{-2}$. Guided by the presence of multiple narrow dips in our spin folded lightcurve, which possibly indicating presence of inhomogeneos absorber, we applied an extra partial covering absorber model, implemented by \texttt{partcov*phabs} on top of the overall absorption. This readily improved the fit to a significant amount ($\chi^2(DOF)=464(397)$ i.e $\Delta\chi^2=52$ for 2 less DOF) with a column density of partial absorber, $n_{\text{H,pcf}}=10.7_{-2.5}^{+3.6}\times10^{22}\; \text{cm}^{-2}$ and a covering fraction of $0.27_{-0.05}^{+0.07}$. However, the model now define the lower energy part better, with the high temperature plasma emission component detecting a smaller value $10.5_{-2.5}^{+3.6}$ keV. The model underestimates the observed data in harder X-ray (beyond 7 keV), resulting in excess in residual. This motivated us to add one more plasma emission model, which produced an improved fit-staistic of $\chi^2(DOF)=429(394)$ with maximum temperature of $36.2_{-13.1}^{+32.9}\; \text{keV}$. F-test probability corresponding to the third \texttt{mekal} component is $8.75\times10^{-7}$, signifying its necessity. Though this temperature is not very well constrained due to absence of extended hard X-ray data for XMM Newton EPIC, yet clearly accounts for the excess residual beyond 7 keV, and is in agreement with the upper temperature we obtain from NuSTAR-only fit.

\begin{figure}
	\includegraphics[width=\columnwidth, trim = {0.5cm 0.4cm 3.6cm 2.7cm}, clip]{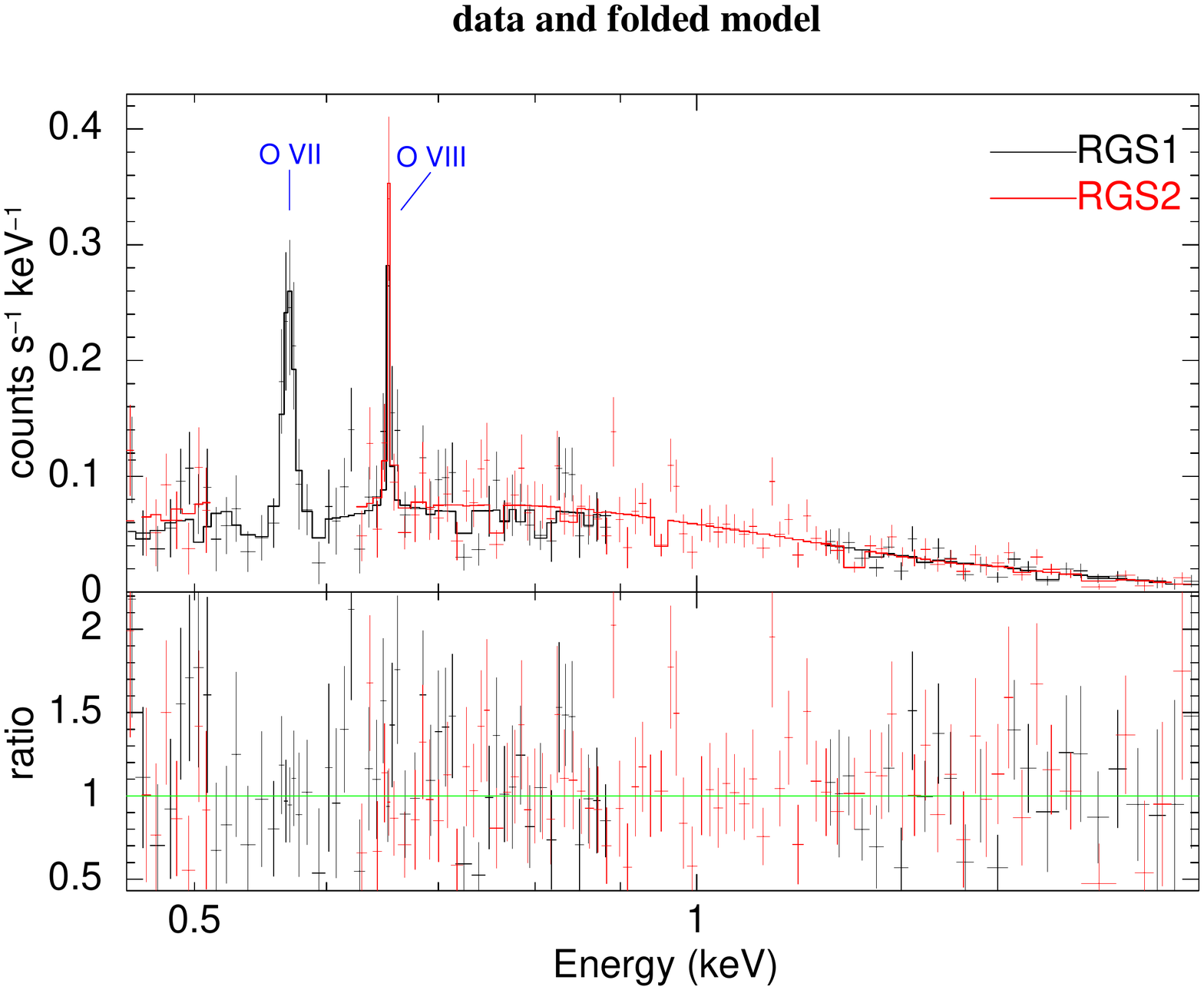}
    \caption{Modelling the ionised oxygen lines (O VII and O VIII) using an absorbed bremsstrahlung continuum along with two gaussian component in 0.45-2.0 keV. The \textcolor{blue}{top} panel shows the spectrum, and the \textcolor{blue}{bottom} panel shows ratio (data/model) plot.}
    \label{fig:oxygenfig}
\end{figure}

We have also looked into the grating spectrometer data, RGS in 0.45-2.0 keV band. We have used an absorbed thermal bremsstrahlung continuum model for fitting the data, with column density of absorber as a free quantity. The upper limit of the continuum temperature became unbounded, so kept fixed at fit value $4.95$ keV. We see clear line emissions at $\sim$ 0.57 keV and 0.65 keV, corresponding to the ionised oxygen K$_{\alpha}$ emission lines (O VII and O VIII respectively) in the spectra. We added two gaussian components to model those emission features. This produced a fit statistic of $\chi^2(DOF)=161(163)$. The corresponding best fit parameter values are listed in  Table~\ref{tab:oxygentab} with the best fit spectrum plot in Fig~\ref{fig:oxygenfig}. We noticed that the  O VIII line is narrow, for which the width ($\sigma$) parameter couldn't be constrained and reaching value lower than instrument resolution, so we fixed it to 0. The width ($\sigma$) of O VII line was allowed to vary, however its best fit value reached almost the instrument resolution limit. The O VII line is composed of fine atomic transition lines (resonance, intercombination and forbidden lines), which are not resolved in the RGS spectra. The clear appearance of ionised oxygen K$_{\alpha}$ lines agrees with the presence of strong excess around 0.5-0.7 keV in EPIC spectra. Each of the gaussian components improve the fit w.r.t absorbed continuum by $\Delta\chi^2 \sim 31$ for 2 less DOF, indicating strong statistical significance of the two lines. The grating spectra doesn't show presence of other such strong lines, so modelling them using gaussian components are not statistically significant.

\begin{table}
	\centering
	\caption{Best fit parameters from fitting oxygen K$_{\alpha}$ lines (0.45-2.0 keV RGS data)}
	\label{tab:oxygentab}
	\renewcommand{\arraystretch}{1.4}
\begin{tabular}{ccc}
\hline
Parameter          						& Unit                  & Value                     \\ \hline
n$_{H}^{\dagger\;a}$            		& $10^{20}\;\; cm^{-2}$ & $4.68_{-1.89}^{+1.97}$    \\ \hline
T$_{C}^{\dagger\;b}$            		& keV                   & $4.95_{\mathit{fr}}$   \\
N$_{C}^{\dagger\;c}$					& $10^{-3}$             & $1.44_{-0.12}^{+0.12}$    \\ \hline
LineE$_{\text{He-like}}^{\dagger\;d}$  	& keV                   & $0.569_{-0.002}^{+0.002}$ \\
$\sigma_{\text{He-like}}^{\dagger\;e}$ 	& eV                    & $3.9_{-1.3}^{+1.7}$       \\
eqw$_{\text{He-like}}^{\dagger\;f}$    	& eV                    & $42_{-15}^{+15}$          \\
N$_{\text{He-like}}^{\dagger\;g}$		& $10^{-5}$             & $10.28_{-3.56}^{+4.39}$  \\ \hline
LineE$_{\text{H-like}}^{\dagger\;d}$   	& keV                   & $0.654_{-0.001}^{+0.008}$ \\
$\sigma_{\text{H-like}}^{\dagger\;e}$  	& eV                    & $0_{{\text{fr}}}$      \\
eqw$_{\text{H-like}}^{\dagger\;f}$     	& eV                    & $17_{-5}^{+5}$            \\
N$_{\text{H-like}}^{\dagger\;g}$		& $10^{-5}$             & $3.51_{-1.09}^{+1.34}$    \\ \hline
$\chi^{2}(DOF)$     					&                       & 161(163)                  \\ \hline
$\chi^2_{\nu}$   						&                       & 0.9877                    \\ \hline
\end{tabular}

\begin{tablenotes}
      \small
      \item $\dagger\;a$ : Overall column density
	  \item $\dagger\;b, c$ : Temperature and normalisation of bremsstrahlung continuum 
	  \item $\dagger\;d, e, f, g$ : Line energy, $\sigma$, equivalent width, and normalisation (in terms of photons cm$^{-2}$ s$^{-1}$) of the corresponding gaussian component respectively
	  \item * $\mathit{fr}$ denotes the parameter is fixed.
\end{tablenotes}	  

\end{table}

This phenomenological fit of XMM-Newton EPIC data now guides us to construct the final model in the next subsection for broadband spectral analysis of simultaneous data, obtained from both the observatories.

\subsubsection{Broadband spectra fitting using XMM-Newton EPIC and NuSTAR}

\begin{figure}
	\includegraphics[width=\columnwidth, trim = {0.5cm 0.4cm 3.6cm 2.7cm}, clip]{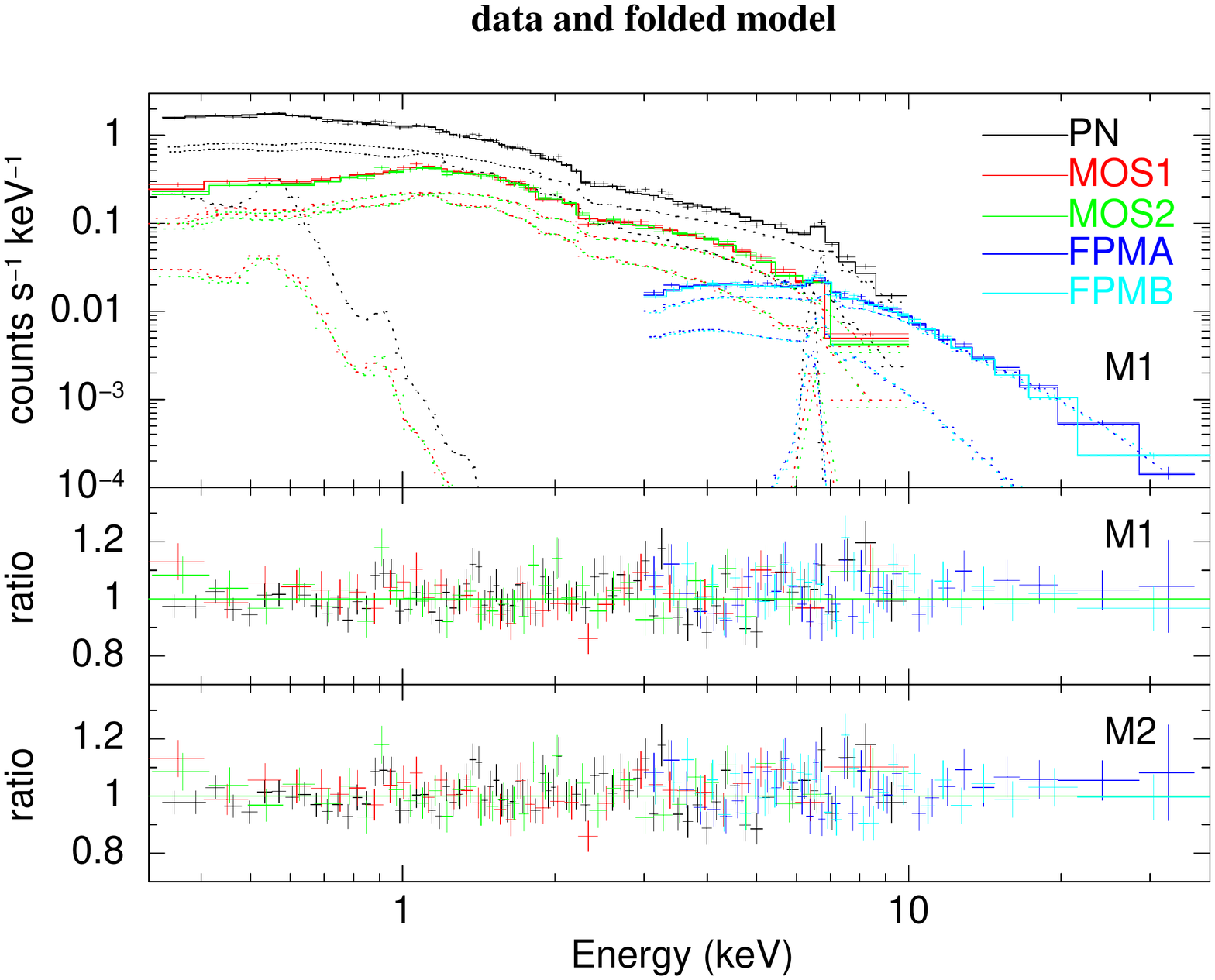}
    \caption{Best fit broadband spectra in 0.3-40 keV with model M1 is shown in \textcolor{blue}{top} panel. The ratio (data/model) plot for spectral fitting with model M1  and M2 is shown in \textcolor{blue}{middle} and \textcolor{blue}{bottom} panel respectively}.
    \label{fig:fitfig}
\end{figure}

We have used the absorbed multi-temperature hot plasma emission model, as developed during phenomenological fits, for modelling the broadband data in 0.3-40.0 keV range. We used model \texttt{cons*tbabs*phabs*(partcov*phabs)* (mekal+mekal+mekal+gauss)} (model M1) to fit the broadband data, producing a resultant fit statistic $\chi^{2}(DOF)=839(857)$. The best fit parameters are quoted in Table \ref{tab:fittab} with the spectra shown in Figure \ref{fig:fitfig}. The fit can describe spectra perfectly by incorporating a total and a partial covering absorber, signifying complex absorption and with clearly detected atleast three plasma temperatures. Best-fit parameters agree with phenomenological fits, but with better error constraints. 

We have also tried adding one more plasma emission component (\texttt{mekal}), resulting in a marginal improvement of fit statistic ($\chi^2(DOF)=835(855)$). The corresponding F-statistic probability is 0.094.
%unlike F-test probability of $5.11799\times10^{-19}$ for the inclusion of third \texttt{mekal} component.
So, we didn't find a strong incentive to keep this extra fourth component. 

At this stage our spectral modelling of broadband data constrains plasma emission parameters as well as column density parameters of absorption components. Next, in order to check the effect of Compton reflection, we have convolved the \texttt{reflect} model with the plasma emission components using the broadband data. The abundance parameters in \texttt{mekal} and \texttt{reflect} were linked and kept free. Similar to NuSTAR-only fit, there is no improvement to fit statistics with negligibly small reflection amplitude ($\sim10^{-5}$).

To check the robustness of the parameters obtained from our simple yet effective three temperature plasma emission model (M1), we have also modelled our spectra using isobaric cooling flow model for ionised plasma, \texttt{mkcflow} \citep{Mushotzky_1988, Mukai_2003}. This model considers multi temperature nature of spectra by using the emissivity function as inverse of bolometric luminosity. The redshift parameter of \texttt{mkcflow} was kept fixed at $5.53\times10^{-8}$ according to the GAIA DR3 distance of $237\pm4$pc \citep{GAIA_2021}. The switch parameter was set at 2. We have used model \texttt{cons*tbabs*phabs*(partcov*phabs)* (mkcflow+mekal+gauss)} (model M2) for the broadband fit ($\chi^2(DOF)=839(858)$) and best-fit parameters are quoted in Table ~\ref{tab:fittab} with ratio plot in Fig. \ref{fig:fitfig}. We noticed the absorption parameter values agree with model M1, thus independent of choice of the model. We required the extra optically thin plasma emission component to consider the excess around 0.6 keV. Fixing low temperature of cooling flow component at 0.0808 keV could not consider the excess around 0.6 keV and actually gave poorer fit statistic. So, we kept the lower temperature parameter of cooling flow component free, which detects some mean temperature near the base of the PSR where matter cools rapidly. The upper temperature is $43.3_{-3.4}^{+3.8}$ keV, which is relatively higher than model M1 ($26.3^{+5.3}_{-3.4}$ keV). Nevertheless, the fit statistics for both M1 and M2 models agree well. At this point, we included Compton reflection to test the necessity of it in our fit with M2 model, and observed similar findings as that of M1, signifying that reflection is not demanded by present data.

\begin{table*}
	\centering
	\caption{Best-fit parameters obtained from fitting simultaneous broadband (0.3-40.0 keV) data}
	\label{tab:fittab}
	\renewcommand{\arraystretch}{1.4}

\begin{tabular}{cccccccc}
\hline
                   &                          & \multicolumn{2}{c}{Spin-Phase-Averaged Spectra} & \multicolumn{4}{c}{Spin-Phase-Resolved Spectra during Phase}                                         \\ \hline
                   &                          &                        &                        & \multicolumn{2}{c}{0.05-0.75}                    & \multicolumn{2}{c}{0.75-1.05}                     \\ \cline{5-8} 
Parameters         & Unit                     & M1                     & M2                     & M1                      & M2                     & M1                      & M2                      \\ \hline
$n_{\text{H,tb}}$  & $\times10^{19}\;cm^{-2}$ & $1.0_{\mathit{fr}}$    & $1.0_{\mathit{fr}}$    & $1.0_{\mathit{fr}}$     & $1.0_{\mathit{fr}}$    & $1.0_{\mathit{fr}}$     & $1.0_{\mathit{fr}}$     \\
$n_{\text{H,ph}}$  & $\times10^{20}\;cm^{-2}$ & $1.84_{-0.70}^{+0.99}$ & $1.80_{-0.69}^{+0.84}$ & $1.23_{-0.68}^{+0.69}$  & $1.37_{-0.63}^{+0.60}$ & $8.78_{-3.45}^{+2.96}$  & $9.74_{-2.97}^{+2.61}$  \\
$n_{\text{H,pcf}}$ & $\times10^{22}\;cm^{-2}$ & $7.44_{-1.76}^{+2.39}$ & $6.84_{-1.68}^{+2.21}$  & $11.22_{-3.22}^{+4.72}$ & $9.60_{-3.55}^{+3.95}$ & $5.26_{-2.01}^{+2.95}$  & $6.68_{-2.37}^{+3.48}$ \\
cvr frac           &                          & $0.25_{-0.04}^{+0.04}$ & $0.26_{-0.04}^{+0.05}$  & $0.23_{-0.03}^{+0.05}$  & $0.26_{-0.05}^{+0.03}$ & $0.33_{-0.07}^{+0.07}$  & $0.32_{-0.09}^{+0.06}$  \\
$T_1$              & keV                      & $0.15_{-0.04}^{+0.03}$ & $0.15_{-0.04}^{+0.03}$ & $0.17_{-0.03}^{+0.02}$  & $0.17_{-0.05}^{+0.03}$  & $<0.10$ & $<0.10$   \\
$T_2$              & keV                      & $4.9_{-1.2}^{+1.1}$    & $1.9_{-0.5}^{+0.6}$    & $4.4_{-0.9}^{+1.4}$     & $1.5_{-0.5}^{+0.6}$    & $4.0_{-0.9}^{+1.1}$     & $2.3_{-0.9}^{+2.9}$     \\
$T_3$              & keV                      & $26.2_{-3.4}^{+5.3}$   & $43.3_{-3.4}^{+3.8}$   & $22.2_{-3.1}^{+5.3}$    & $39.4_{-3.3}^{+5.1}$   & $26.2_{\mathit{fr}}$    & $43.3_{\mathit{fr}}$    \\
$Z$                & $Z_{\odot}$              & $1.14_{-0.19}^{+0.22}$  & $1.11_{-0.02}^{+0.02}$ & $1.08_{-0.20}^{+0.26}$  & $0.98_{-0.19}^{+0.19}$ & $1.03_{-0.35}^{+0.43}$  & $0.82_{-0.29}^{+0.33}$  \\
LineE              & keV                      & $6.43_{-0.03}^{+0.03}$ & $6.44_{-0.03}^{+0.03}$  & $6.44_{-0.08}^{+0.03}$  & $6.44_{-0.04}^{+0.05}$ & $6.46_{-0.06}^{+0.07}$  & $6.48_{-0.06}^{+0.08}$   \\
$\sigma$           & eV                       & $0_{\mathit{fr}}$      & $0_{\mathit{fr}}$      & $0_{\mathit{fr}}$       & $0_{\mathit{fr}}$      & $0_{\mathit{fr}}$       & $0_{\mathit{fr}}$       \\
$N_L$              & $\times10^{-5}$          & $0.95_{-0.25}^{+0.29}$ & $1.04_{-0.22}^{+0.26}$ & $0.85_{-0.34}^{+0.30}$  & $0.88_{-0.28}^{+0.32}$  & $1.19_{-0.52}^{+0.51}$   & $1.18_{-0.50}^{+0.50}$  \\ \hline
$\chi^2(DOF)$      &                          & 839(857)               & 839(858)               & 694(752)                & 688(753)               & 429(427)                & 438(428)                \\
$\chi^2_{\nu}$     &                          & 0.9790                 & 0.9778                 & 0.9229                  & 0.9137                 & 1.005                   & 1.023                   \\ \hline
\end{tabular}

\begin{tablenotes}
      \small
      \item \# Model M1: \texttt{constant*tbabs*phabs*(partcov*phabs)*(mekal+mekal+mekal+gauss)}
      \item \# Model M2: \texttt{constant*tbabs*phabs*(partcov*phabs)*(mkcflow+mekal+gauss)}
      \item \# For model \textbf{M1}: $T_1$, $T_2$, $T_3$ denotes temperature of the three plasma emission components (\texttt{mekal}) % and "N" denotes normalisation of corresponding plasma emission components. 
      \item \# For model \textbf{M2}: $T_1$ denotes temperature of single temperature plasma emission \texttt{mekal} component; %and $N_1$ denotes corresponding normalisation. 
      $T_2, T_3$ denotes upper and lower temperature of cooling flow component \texttt{mkcflow} %$N_3$ denotes corresponding normalisation.
      \item \# $n_{\text{H,tb}}$, $n_{\text{H,ph}}$, $n_{\text{H,pcf}}$ denotes equivalent column density of the ISM, overall absorber and the partial covering absorber respectively. "cvr frac" stands for the covering fraction of the partial covering absorber.
      \item \# $Z$ denotes the overall abundance
      \item \# LineE, $\sigma$, $N_L$ represents line energy, $\sigma$ and normalisation (in terms of photons cm$^{-2}$ s$^{-1}$) of the gaussian component for neutral Fe K$\alpha$ line.
      \item \# $\mathit{fr}$ denotes the parameter is fixed.

\end{tablenotes}

\end{table*}

\subsubsection{Phase-resolved spectroscopy}

\begin{figure}
	\includegraphics[width=\columnwidth, trim = {0.5cm 0.4cm 3.6cm 2.5cm}, clip]{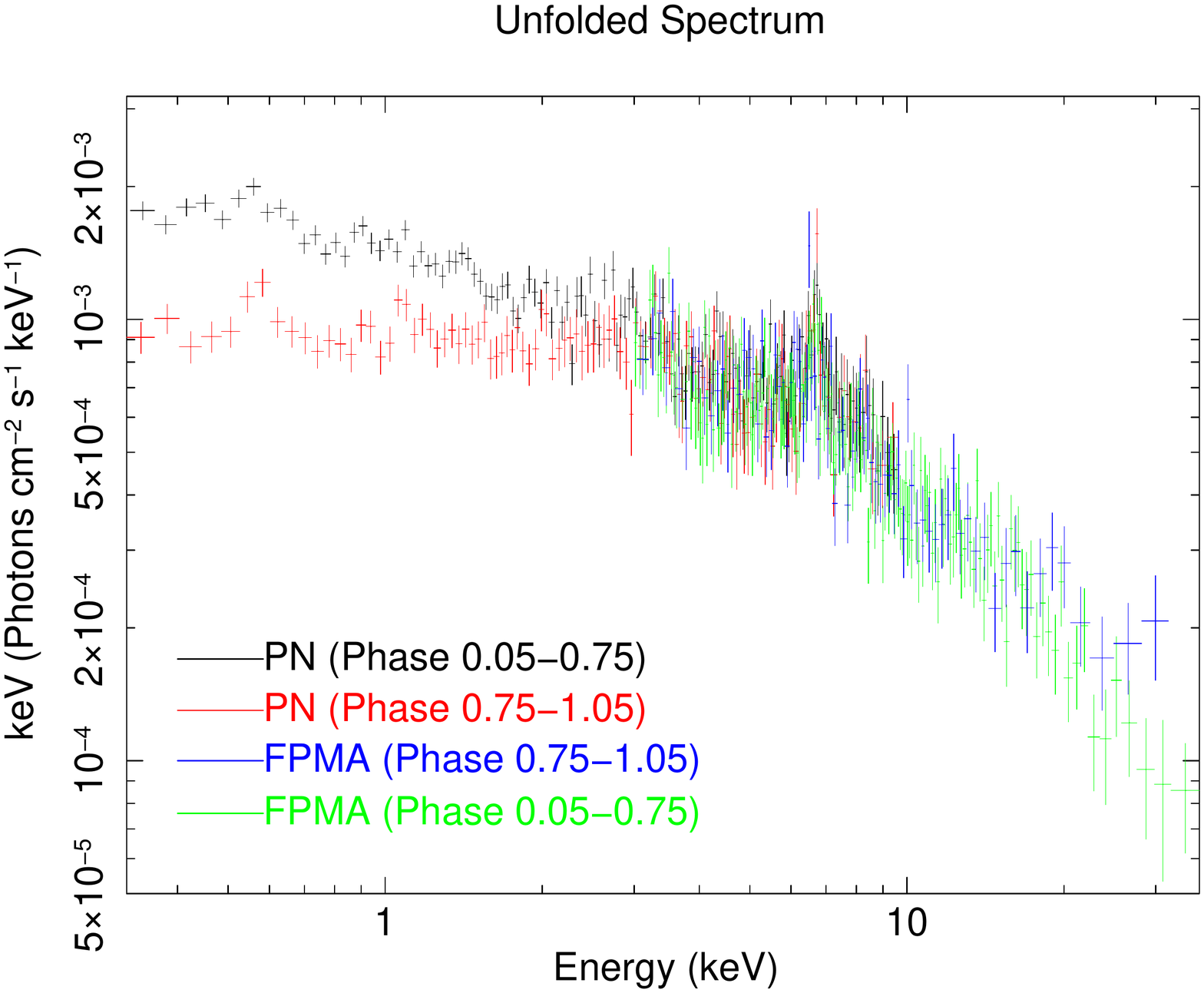}

    \caption{ Comparison of phase-resolved spectra in 0.05-0.75 and 0.75-1.05 spin phase. An increased absorption below 3 keV is prominent for spin phase 0.75-1.05.} 
    \label{fig:specphase}
\end{figure}

The HR1 (bottom panel of Fig. \ref{fig:spinfoldPN}) shows excess during the spin phase $\sim0.75-1.05$ indicating spectral variation. However, HR2 (bottom panel of Figure \ref{fig:spinfoldFPMA}) doesn't indicate any such strong variation during the entire spin phase. The spin phase-resolved spectra in Figure \ref{fig:specphase}, noticeably exhibit that the lower energy part of the spectra (below $\sim3\;\text{keV}$) is more absorbed during the phase 0.75-1.05 compared to phase 0.05-0.75, thereby explaining the excess in HR1 plot. To evaluate the spectral parameters, we have considered the broadband phase-resolved spectra in 0.3-40.0 keV range, and fit with the same models as phase-averaged spectra. The obtained best fit parameters are tabulated in the Table ~\ref{tab:fittab}. Using the spectral model M1, the equivalent column density of overall photoelectric absorber comes out to be $8.78_{-3.45}^{+2.96}\times 10^{20}\;\text{cm}^{-2}$ during phase 0.75-1.05, much higher than $1.23_{-0.68}^{+0.69}\times 10^{20}\;\text{cm}^{-2}$  during phase 0.05-0.75. The best fit values of equivalent column density of partial covering absorber and covering fraction for the phase 0.05-0.75 are $11.2_{-3.2}^{+4.7}\times 10^{22}\;\text{cm}^{-2}$ and $0.23_{-0.03}^{+0.05}$ respectively. The same parameters for the phase 0.75-1.05 have values $5.3_{-2.0}^{+2.9} \times 10^{22}\;\text{cm}^{-2}$ and $0.33 \pm 0.07$ respectively. The above mentioned two parameters have slightly different best fit values but with overlapping error bars. Similar parameter values are obtained with model M2. The temperature of cold plasma emission component has an upper limit of $\sim 0.1$  keV in phase 0.75-1.05, but the lower limit is unconstrained (reaching the minimum temperature $ \sim 0.0808$ keV, allowed by the model). This temperature could be mimicking the temperature near the bottom of the PSR. The upper temperature was getting poorly constrained in the phase 0.75-1.05, because of the reduced count statistics of NuSTAR data in that phase. So, we kept that temperature fixed at the corresponding value of best fit temperature obtained from phase-averaged spectra.

\section{Discussion}
\label{sec:discussion}

We have carried out the broadband X-ray timing and spectral analysis of the asynchronous polar CD Ind. The implication of the timing and spectral results are discussed in this section.

\subsection{Spin modulations in folded lightcurves}

The spin folded lightcurve in 0.3-3.0 keV band (Fig ~\ref{fig:spinfoldPN}) shows strong pulse profile with a single broad hump with multiple narrow dips and a pulse fraction of $62\pm2\%$. The single broad hump picture nicely fits into the earlier reporting of one pole accretion at any specific beat phase \citep{Littlefield_2019, Myers_2017, Ramsay_2000}. The dips represent the scenario where the X-ray emission passes through the inhomogeneous and complex accretion stream, undergoing photoelectric absorption. The minima at phase $\sim0.94$ can be envisaged as when the corresponding emitting region is moving away from the line of sight and the emission is reaching us after passing through the intrinsic absorber. The broad hump like structure is missing in 3.0-10.0 keV band of PN, but the features of absorption dips are present (Fig ~\ref{fig:spinfoldPN}). It indicates the corresponding X-ray emitting region of the PSR remains visible during the entire spin cycle, but still passes through the inhomogeneous accretion stream. Also, the spin folded NuSTAR lightcurve in the 3.0-10.0 keV band, as well as 10-40 keV band (Fig ~\ref{fig:spinfoldFPMA}) lack strong modulation.
%\sout{This is expected from the NuSTAR power spectrum where the peak corresponding to spin period is relatively weaker.}
The lack of strong modulation in hard X-rays suggests corresponding zones of the PSR remain in the view throughout the spin phase. This scenario is representative of a tall PSR and a small angle between spin axis and magnetic axis (eg. $\sim 10^{\circ}$, \citet{Ramsay_2000}).

\subsection{Multi-temperature nature of the accretion column}

Our spectral analysis of the spin-average broadband spectra reveals the multi-temperature nature of the post-shock plasma. It is represented by three plasma temperatures in our spectral modelling. The upper temperature is close to the shock temperature. The lower temperature, coming from the optically thin cold plasma, indicates the scenario near the bottom of the PSR. The middle-temperature plasma designates the cumulative contribution from  remaining temperature zones of the PSR. It is to be mentioned that in Aizu model, PSR consists of multiple temperature zones with gradually decreasing temperature with most of cooling occurring near bottom of the PSR (Fig. 2 of \citet{Mukai_2017}), and there is no distinct division among them. Our three-component plasma emission model is a simple yet useful picture to represent the PSR, with averaged contribution of plasma emissivity from different zones. A more extensive approach to represent the multi-temperature nature of PSR is by cooling flow models \citep{Mukai_2003}. However, the lower temperature of cooling flow model in our fit does not necessarily indicate the lowest temperature at bottom of the PSR and possibly detects some mean value of temperature near the base of the PSR where the plasma is rapidly cooling. The requirement of an additional optically thin plasma to incorporate the excess due to emission features present in soft X-ray is also evident in cooling flow model.

The importance of having highly sensitive spectral data in hard X-rays is to measure the shock temperature accurately, thereby constraining the mass of the WD. The upper temperature of the cooling flow model represents the shock temperature more accurately than the three-component plasma emission model. This is because, cooling flow model considers the temperature gradient of the PSR using multiple grid points according to emissivity function of the emitting plasma, whereas the three-component plasma emission model represents the whole PSR using three plasma temperatures. Thus the upper temperature for three component plasma emission model is more likely to be an average value of temperature profile from the region below the shock. Using the relation between the shock temperature, mass and radius of the WD \citep{Mukai_2017}, and incorporating the WD mass-radius relation \citep{Nauenberg_1972}, we quote a mass value $M_{\mathit{WD}}=0.87_{-0.03}^{+0.04} M_{\odot}$ for a shock temperature of $43.3_{-3.4}^{+3.8}\; \text{keV}$ (obtained from model M2, see Table \ref{tab:fittab}). The corresponding radius is $R_{\mathit{WD}}=6.42_{-0.28}^{+0.27} \times 10^8 \; \text{cm}$ ($\sim0.009\;R_{\odot}$). \citet{Ramsay_2000} measured a WD mass of $M_{\mathit{WD}}=0.79_{-0.11}^{+0.12} M_{\odot}$ using RXTE PCA spectra in 4-15 keV. Our measured mass, obtained using broadband spectra, matches with them, but with better error constraint.

The line of sight orbital velocity of the WD comes out to be $\sim90\;\text{km}/s$, calculated using the following parameters: the mass of WD (obtained in this work $\sim 0.87 M_{\odot}$), mass of M6V secondary (\cite{Littlefield_2019}, typical mass $\sim 0.21 M_{\odot}$), binary period $\sim 6720$ s \citep {Littlefield_2019}, and inclination angle of $\sim 65^{\circ}$ \citep{Mason_2020}.

We obtain an unabsorbed bolometric (0.3-40.0 keV) flux of $22.2_{-0.4}^{+0.4}\times10^{-12}\;\text{erg}\;\text{cm}^{-2}\;\text{s}^{-1}$ of which $\sim64\%$ is contributed from 0.3-10.0 keV band and the remaining is from 10-40 keV band. The corresponding luminosity from the source is $L=1.49_{-0.07}^{+0.08}\times10^{32} \;\text{erg}\;\text{s}^{-1}$ (using $L=4 \pi F d^{2}$ where $d$ is the distance to the source $\sim 237 \pm 4 \; \text{pc}$ \citep{GAIA_2021}).

The relation between accretion luminosity ($L_{\mathit{acc}}$), mass, radius and mass accretion rate of the WD ($M_{\mathit{WD}}$, $R_{\mathit{WD}}$ and $\dot{M}$ respectively) is given by \citep{Frank_2002},

\begin{center}

$L_{\mathit{acc}}=\frac{GM_{\mathit{WD}}\dot{M}}{R_{\mathit{WD}}}$

\end{center}

Assuming accretion luminosity is mostly emitted in X-rays, we calculate a mass accretion rate of $\dot{M}\sim8.24\times10^{14}\;\text{g}\,\text{s}^{-1}\:\sim1.30\times{10^{-11}}\;M_{\odot}\,\text{yr}^{-1}$ using our obtained values of mass, radius and luminosity. 

\subsection{Iron and Oxygen K$_{\alpha}$ line emissions}
\label{sec:ironreflect}
 
The asynchronous polar CD Ind showed strong Fe K$_{\alpha}$ line emission, and XMM-Newton could resolve the three lines i.e. fluorescence, He-like and H-like lines. The line diagnostic shows that all the three Fe K$_{\alpha}$ lines in CD Ind have central energies at their expected positions within 90\% confidence level, and the lines are narrow, which is expected from the radial velocity of the emitting pole of the WD. The intensity and equivalent width of the He-like Fe line are the strongest among the three lines. On the other hand, the neutral Fe K$_{\alpha}$ line is weakest among all, carrying an equivalent width of $71_{-30}^{+37}\;eV$.

According to \cite{EzukaIshida_1999}, the observed equivalent width of the neutral Fe-line could be a sum of contribution from various components, like absorbing material and the cold surface of the WD. However, within the current limitation of the data, it is not possible to distinguish between these contributions, therefore we observe a total equivalent width.
The Compton reflection is also originated from the similar region of the WD when the hard X-ray emission hits cold material at surface \citep{van_teeseling_1996}. Our data could not unambiguously detect the presence of reflection, which might be very small in the spectra. If the shock height is very large, the emitted hard X-rays can subtend only a very small solid angle to the WD surface, thereby producing a negligible reflection.

In addition to the Fe K$_{\alpha}$ line in EPIC spectra, the RGS grating spectra show presence of strong ionised oxygen K$_{\alpha}$ lines, appearing with equivalent width of $42\pm15$ eV (O VII) and $17\pm5$ eV (O VIII). These lines come from the cooler bottom region of the PSR. Presence of these lines with such strength indicates that the bottom temperature of the PSR is low enough to produce them. This fits with our prediction about tall shock height in CD Ind, so that the PSR gets sufficient time to cool down while reaching WD photosphere.

Previous study of CD Ind from XMM-Newton also reported presence of a strong Ni K$_{\alpha}$ line at 7.4 keV \citep{Joshi_2019}. However, we did not find any such emission feature in our XMM-Newton data.

\subsection{Excess absorption during spin phase 0.75-1.05}

Our phase-resolved spectroscopy of CD Ind clearly identified the increased absorption in the soft X-rays below 3 keV during 0.75-1.05 spin phase as indicated by the HR1 of spin folded lightcurves. The column density of overall photoelectric absorption, which affects the low energy part increased by almost an order of magnitude. The partial covering absorber, however, did not change significantly. Due to more absorption, the absorbed flux value obtained in 0.3-3.0 keV band during phase 0.75-1.05  ($3.89_{-0.18}^{+0.15}\times10^{-12}\;\text{erg}\;\text{cm}^{-2}\;\text{s}^{-1}$) is less than the flux in same band during phase 0.05-0.75 ($5.50_{-0.09}^{+0.06}\times10^{-12}\;\text{erg}\;\text{cm}^{-2}\;\text{s}^{-1}$). The absorbed flux values in 3-40 keV band during both phases remained similar.

\section{Summary}
\label{sec:summary}

Our study using simultaneous broadband X-ray data has enabled us to extend the understanding of the accretion properties of the asynchronous polar CD Ind. Here, we summarise our results:

\begin{itemize}

\item
Our X-ray observation supports single-pole accretion model where one accretion region is active and remains visible throughout the spin phase. This fits with the existing picture of pole-switching scenario of the source where alternatively two poles become active during two different phases of a beat cycle.

\item
Presence of complex absorber is indicated in our study. The emitted X-rays pass through the inhomogeneous accretion stream, causing several narrow dips in the folded lightcurve. We also notice a significant increase in column density of the overall absorber for certain spin phase, affecting the soft X-rays below $\sim 3$ keV.

\item
We constrained the mass of the WD to be $0.87^{+0.04}_{-0.03}\; M_{\odot}$. This is directly measured from the shock temperature using the extended hard X-ray data from NuSTAR, and thus an improvement from the earlier measured masses.

\item
We could not unambiguously detect the Compton reflection, which may be small and might not have revealed itself in our spectra. We predict a possible scenario where the shock height is large. 

\item
The bottom of the PSR cools down sufficiently, as supported by the presence of strong ionised oxygen K$_{\alpha}$ lines in the spectra.

\end{itemize}

In this X-ray broadband study of  CD Ind, there are certain limitations. We could only look into a part of the beat phase using our joint simultaneous data. Also our observation from XMM-Newton in 0.3-10.0 keV only covers a small subset of NuSTAR observation which includes 10-40 keV. \citet{Ramsay_2000} didn't find significant difference in X-ray spectra between two different phases of a beat cycle, when two different poles were acreting, but their observation was based on limited 4-15 keV band of RXTE-PCA. Considering the complex variability of accretion mechanism in the source (as predicted in recent works eg. \cite{Mason_2020, Sobolev_2021}), a future monitoring X-ray campaign for simultaneous broadband data covering several beat phases, and possibly in several beat cycles (for improved count statistics in hard X-rays) will bring out a clearer impression about the change in accretion properties over a complete beat phase.

%%\label{sec:maths} % used for referring to this section from elsewhere

%Simple mathematics can be inserted into the flow of the text e.g. $2\times3=6$
%or $v=220$\,km\,s$^{-1}$, but more complicated expressions should be entered
%as a numbered equation:

%\begin{equation}
%    x=\frac{-b\pm\sqrt{b^2-4ac}}{2a}.
%	\label{eq:quadratic}
%\end{equation}

%Refer back to them as e.g. equation~(\ref{eq:quadratic}).

%\subsection{Figures and tables}

%Figures and tables should be placed at logical positions in the text. Don't
%worry about the exact layout, which will be handled by the publishers.

%Figures are referred to as e.g. Fig.~\ref{fig:example_figure}, and tables as
%e.g. Table~\ref{tab:example_table}.

%% Example figure
%\begin{figure}
%	% To include a figure from a file named example.*
%	% Allowable file formats are eps or ps if compiling using latex
%	% or pdf, png, jpg if compiling using pdflatex
%	\includegraphics[width=\columnwidth]{example}
%    \caption{This is an example figure. Captions appear below each figure.
%	Give enough detail for the reader to understand what they're looking at,
%	but leave detailed discussion to the main body of the text.}
%    \label{fig:example_figure}
%\end{figure}

%% Example table
%\begin{table}
%	\centering
%	\caption{This is an example table. Captions appear above each table.
%	Remember to define the quantities, symbols and units used.}
%	\label{tab:example_table}
%	\begin{tabular}{lccr} % four columns, alignment for each
%		\hline
%		A & B & C & D\\
%		\hline
%		1 & 2 & 3 & 4\\
%		2 & 4 & 6 & 8\\
%		3 & 5 & 7 & 9\\
%		\hline
%	\end{tabular}
%\end{table}

\section*{Acknowledgements}

We would like to thank the referee for positive comments that helped in further improving the manuscript. This research has made use of the data obtained from NuSTAR telescope, operated jointly by CalTech and NASA, and XMM-Newton telescope, operated by ESA. We thank the NuSTAR science operation team and XMM-Newton science operation team for providing the data. We acknowledge the members at the helpdesk, maintained by High Energy Astrophysics Science Archive Research Center (HEASARC) and the XMM-Newton helpdesk team members for providing necessary support. A.D mentions other useful help provided by H.M, T.G and S.K during the work.
%The Acknowledgements section is not numbered. Here you can thank helpful colleagues, acknowledge funding agencies, telescopes and facilities used etc. Try to keep it short.

%%%%%%%%%%%%%%%%%%%%%%%%%%%%%%%%%%%%%%%%%%%%%%%%%%
\section*{Data Availability}

The data used for analysis in this article are publicly available in NASA's High Energy Astrophysics Science Archive Research Center (HEASARC) archive (\url{https://heasarc.gsfc.nasa.gov/docs/archive.html}) and XMM-Newton Science archive (\url{http://nxsa.esac.esa.int/nxsa-web/#search}). The observation IDs are mentioned in Sect.~\ref{sec:ObsDataRed}.

%The inclusion of a Data Availability Statement is a requirement for articles published in MNRAS. Data Availability Statements provide a standardised format for readers to understand the availability of data underlying the research results described in the article. The statement may refer to original data generated in the course of the study or to third-party data analysed in the article. The statement should describe and provide means of access, where possible, by linking to the data or providing the required accession numbers for the relevant databases or DOIs.



\bibliographystyle{mnras}
\bibliography{mnras_template_cdind} % if your bibtex file is called example.bib

%\begin{thebibliography}{99}
%\end{thebibliography}

%%%%%%%%%%%%%%%%%%%%%%%%%%%%%%%%%%%%%%%%%%%%%%%%%%

%%%%%%%%%%%%%%%%% APPENDICES %%%%%%%%%%%%%%%%%%%%%

%\appendix



%%%%%%%%%%%%%%%%%%%%%%%%%%%%%%%%%%%%%%%%%%%%%%%%%%

% Don't change these lines
\bsp	% typesetting comment
\label{lastpage}
\end{document}